\documentclass[trackchanges]{aastex701}

\begin{document}

\title{A high-frequency type II radio burst associated with an X2.3 class flare}

\correspondingauthor{Anshu Kumari}
\email{anshu@prl.res.in}

\author[orcid=0009-0001-7689-0084,sname=Paliwal,gname=Divya]{Divya Paliwal}
\affiliation{Udaipur Solar Observatory, Physical Research Laboratory, Dewali, Badi Road, Udaipur-313001, Rajasthan, India}
\affiliation{Indian Institute of Technology Gandhinagar, Gujarat-382355, India
}
\email{divya.paliwal@email.com}

\author[orcid=0000-0001-5742-9033,sname=Kumari,gname=Anshu]{Anshu Kumari}
\affiliation{Udaipur Solar Observatory, Physical Research Laboratory, Dewali, Badi Road, Udaipur-313001, Rajasthan, India}
\email{anshu@prl.res.in}

\author[orcid=0009-0005-3692-1893, sname=Singh,gname=Vishwa]{Vishwa Vijay Singh}
\affiliation{Udaipur Solar Observatory, Physical Research Laboratory, Dewali, Badi Road, Udaipur-313001, Rajasthan, India}
\affiliation{Indian Institute of Technology Gandhinagar, Gujarat-382355, India
}
\email{vishwa.singh@email.com}

\author[orcid=0009-0000-3793-0779, sname=Mishra,gname=Dinesh]{Dinesh Mishra}
\affiliation{Udaipur Solar Observatory, Physical Research Laboratory, Dewali, Badi Road, Udaipur-313001, Rajasthan, India}
\affiliation{Indian Institute of Technology Gandhinagar, Gujarat-382355, India
}
\email{dineshm@prl.res.in}

\author[orcid=0009-0005-9842-709X, sname=Das,gname=Pritam]{Pritam Das}
\affiliation{ Aryabhatta Research Institute of Observational Sciences,  Nainital, Uttarakhand, India}
\email{pritamd9818@gmail.com}

\author[sname=K,gname=Nadiya]{Nadiya K}
\affiliation{Mahatma Gandhi University, Kerala, Kottayam, India}
\email{knadiyavtr@gmail.com}

\shorttitle{A high-frequency type II radio burst associated with an X2.3 class flare}
\shortauthors{Paliwal et al.}

\begin{abstract}

Radio observations provide a powerful diagnostic of the solar corona, enabling investigations of dynamic phenomena associated with solar flares, coronal mass ejections (CMEs), and shock waves. We present a multiwavelength analysis of a rare high-frequency type II radio burst (starting frequency of $\sim$750 MHz and frequency drift rate of $\sim$0.5 MHz s$^{-1}$) associated with the X2.3-class solar flare that occurred on 6 November 2024. A propagating EUV disturbance was observed shortly after the flare onset in the SDO/AIA field of view, while radio spectrographs recorded the type II burst between 13:46 and 13:56 UT over a frequency range of $\sim$750 to 45 MHz. Radio imaging observations from the Nançay Radioheliograph (NRH) show that the radio sources propagate southward during the event. X-ray spectroscopy from HEL1OS onboard Aditya-L1 and imaging observations from STIX onboard Solar Orbiter reveal signatures of efficient non-thermal electron acceleration associated with the flare. An NLFFF extrapolation identifies a pre-eruptive magnetic flux rope in the source region, while white-light coronagraph observations obtained during the event show no detectable large-scale CME. The speed of the erupting flux rope, derived from stereoscopic EUV observations, is consistent, within measurement uncertainties, with the shock speed inferred from the radio dynamic spectrum. Together, these observations suggest that compact flux rope eruptions, even in the absence of a detectable white-light CME, can generate low-coronal shocks capable of producing high-frequency type II radio emission.
\end{abstract}

\keywords{Sun: radio emissions --corona -- flare --accelerating electrons--EUV wave  }

\section{Introduction} \label{sec:intro}

Solar radio emission is an important tool for probing different layers of the solar atmosphere. Since radio waves are sensitive to local plasma conditions, they provide valuable information about physical properties at energy-release sites, such as plasma density, temperature, and magnetic field structure. These emissions also help in understanding how highly energetic electrons are accelerated and propagate from the solar corona into the interplanetary medium \citep[see for a review,][ and the references therein]{2026arXiv260322087K}. In addition, solar radio observations enable the study of both plasma heating and non-thermal electrons acceleration during transient events such as solar flares and coronal mass ejections (CMEs) in the solar atmosphere.
Solar radio bursts are direct signatures of electron acceleration during various transient events in the solar atmosphere. Based on their morphological appearance in dynamic spectra, these bursts are classified mainly into five different types (type I--V), each originating from distinct physical processes \citep{white2007solar}. Type I bursts are characterized by short, narrowband emission in the $80$ -- $200$ MHz frequency range \citep{mugundhan2018spectropolarimetric}. They are typically observed alongside an underlying continuum and are associated with electrons accelerated above active regions following magnetic reconnection. Type II radio bursts result from shock-accelerated electrons and are observed at meter wavelengths \citep{Mann1995}. Type III are commonly associated with solar flares and are produced by electron beams propagating along open or quasi-open magnetic field lines, which generate Langmuir waves through the bump--in--tail instability, leading to intense radio emission \citep{2014RAA....14..773R}. Type IV radio bursts are broadband continuum emissions, broadly classified as moving or stationary \citep{2022SoPh..297...98K}. They are generally attributed to gyrosynchrotron or plasma emission, although resonant transition radiation (RTR) has also been proposed as an alternative incoherent mechanism for some decimetric solar radio continua \citep{Nita2005,Fleishman2005}. In the case of moving events, non-thermal electrons are thought to be trapped in the outward-propagating CME loops \citep{Kumari2021}. In contrast, stationary type IV bursts are associated with non-thermal electrons accelerated at the footpoints of the post-flare arcaded loops after liftoff of plasma  \citep{Liu2018}. Finally, type V radio bursts are relatively rare broadband continuum emissions that typically follow type III bursts and last for several minutes. They were believed to originate from the same population of non-thermal electrons responsible for the preceding type III bursts, although their exact emission mechanism remains only partially understood. 

Type II radio bursts are characterized by slowly drifting emission lane/lanes in dynamic spectra, generated by non-thermal electrons accelerated at quasi-perpendicular shock fronts, which emit from the upstream and downstream regions of the shock.
These bursts often exhibit two or more emission lanes, identified as fundamental and harmonic (F--H) pairs, with additional splitting within each band, called split bands \citep{Smerd1974, nadiya2026multilane}. Rarely, type II bursts also consist of numerous fine structures, called `herringbones'.
Herringbones may occur with/without type II bursts (as backbones) and are characterized by high--to low-frequency drifting lanes and vice versa in the dynamic spectra { \citep{Morosan2021}. Recently, studies have reported that approximately $98 \%$ type II radio bursts are associated with accelerated non-thermal electrons due to CME-driven shocks \citep{Kumari2023b}. The remaining $2\%$ of type II bursts may be due to different scenarios. For example, \citep{Maguire2021} reported that the type II burst was generated by a piston shock driven by the jet in the low corona using the Low Frequency Array (LOFAR) imaging observations. \citet{magdalenic2012flare} showed that the impulsive increase of the pressure in the flare was the source of the shock wave, which gave rise to the type II radio burst. Recently, \citet{2023arXiv230511545M} used extreme--ultraviolet (EUV) observations to show the close association of type II bursts with faint EUV waves propagating through the corona. By comparing the speed of the radio source and EUV wave, \citep{2015ApJ...804...88S} suggest that CME--less type II bursts can be produced by loop-driven shock generated by the rapid expansion of a strongly inclined magnetic loop following magnetic reconnection with rising flux. 

Recent advances in remote sensing have significantly improved understanding of these phenomena through high-resolution observations in space, time, and spectral domains. Instruments such as the Extreme Ultraviolet Imager \citep[EUI;][]{Rochus2020} on board the Solar Orbiter \citep[SolO;] []{2020A&A...642A...1M}, Atmospheric Imaging Assembly \citep[AIA;][]{Lemen2012} on board the Solar Dynamics Observatory \citep[SDO;][]{2012SoPh..275....3P}, the Extreme Ultraviolet Imager \citep[EUVI;][]{Wuelser2004} on board the Solar TErrestrial RElations Observatory  \citep[STEREO;][]{2008stmi.book....5K}, and the newly launched the High Energy L1 Orbiting X--ray Spectrometer \citep[HEL1OS;][]{2025SoPh..300..140N} on board Aditya-L1 \citep{2017CSci..113..610S} provide high resolution observations of the solar atmosphere, enabling detailed investigations of the coronal structures and dynamic associated with solar transient event, including the formation and propagation of type II radio bursts \citep{Ma2011}.

CME-less type II radio bursts are of particular interest because they demonstrate that coronal shocks can form without any detectable large-scale CME in white-light coronagraph observations, challenging the conventional view that coronal shocks are exclusively driven by CMEs. However, the physical mechanism and dynamics of magnetic structure responsible for generating such shocks remain poorly understood. In this study, we investigate high-frequency type II bursts without an observable white-light CME in the Large Angle and Spectrometric Coronagraph \citep[LASCO C2/C3;][]{1995SoPh..162..357B} on board Solar and Heliospheric Observatory \citep[SOHO;][]{Brueckner1995} and the COR1/COR2 coronagraphs of the Sun–Earth Connection Coronal and Heliospheric Investigation \citep[SECCHI;][]{Howard2008} instrument suite on board STEREO--A. Using a multi-instrument approach, we examine the dynamics of the erupting flux rope (FR), the associated EUV signatures, and the origin of the shock near the flaring site that produces the observed radio emission.

This article is organized as follows: Section \ref{sec:data} presents the observational datasets and instruments used in this study. Section \ref{sec:analysis} presents the data analysis and results. In Section \ref{sec:discussion}, we discuss the implications of the findings from this event and provide their physical interpretation. Finally, Section \ref{sec:conclusion} summarizes the main conclusion of this work.

\section{Data} \label{sec:data}
\subsection{Radio observations }

A high-frequency type II was observed on 6 November 2024 with multiple ground-based radio spectrographs. The event is particularly interesting because no clear CME signature was detected at higher altitude in white-light coronagraph observations. The type II was associated with an X2.3 class solar flare originating from NOAA Active region 13883 (S08E14)\footnote{\url{https://www.lmsal.com/solarsoft/latest_events_archive/events_summary/2024/11/06/gev_20241106_1324/index.html}}, which was observed by the Geostationary Operational Environmental Satellite \citep[GOES--16;][]{2022SpWea..2003044D} in geosynchronous orbit around the Earth. The flare started at 13:24:00 UT, peaked at 13:40:00 UT, and ended at 13:46:00 UT, with a total duration of approximately 22 min (Figure \ref{Fig:figure1}, bottom panel). To investigate the temporal and spectral evolution of the radio burst, we used radio observations from three complementary ground-based radio spectrographs: the extended Compact Astronomical Low-cost Low-frequency Instrument for Spectroscopy and Transportable Observatory network \citep[e-callisto;][]{2005SoPh..226..143B}, the Solar Radio Spectrometer (SRS) at San Vito, and the Observation Radio pour FEDOME et l'Étude des Éruptions Solaires \citep[ORFEES;][]{orfees21}.  The e-callisto (Compound Astronomical Low-cost Low-frequency Instrument for Spectroscopy and Transportable Observatory) network is a worldwide array of radio spectrometers that provide nearly continuous 24h monitoring of the sun with a temporal resolution of 0.25 sec and a frequency coverage of 45--870 MHz. The SRS monitors solar radio emissions in the 25-180 MHz range at a 3-second cadence. ORFEES, located at Nançay, France, observes the solar corona between 144--1000 MHz and tracks the Sun from 08:00 to 16:00 UT each day with a temporal resolution of 100 ms. High temporal cadence makes ORFFES suitable for studying rapidly evolving solar radio bursts.
In this analysis, we used the e-callisto data for the 45--70 MHz range, the SRS observations for the 70--144 MHz range, and the ORFEES observations for the 144--1000 MHz range. The combination of these instruments provides continuous spectral coverage from 45--1000 MHz ($\sim 1.55-0.7 R_0$), enabling the evolution of the event over a broad frequency range and across a wide range of coronal heights \citep{newkirk1967structure}. Type II bursts observed at higher frequency from $\sim 750$ MHz to 45 MHz at 13:46 and 13:56 UT. Exhibiting both fundamental and harmonic emission lanes at 300 MHZ and 600 MHz band splitting as illustrated in Figure \ref{Fig:figure1} and detailed in the zoomed-in region in Figure \ref{Fig:figure2}. To determine the location of the radio emission source in the solar atmosphere, we used radio images from the Nançay Radioheliograph \citep[NRH;][]{Kerdraon1997}\footnote{\url{https://secchirh.obspm.fr/spip.php?page=survey&hour=day&survey_type=1&dayofyear=20241106}}. The NRH provides two-dimensional images of the solar radio sources at ten frequencies ranging from 150.9 MHz to 450 MHz with 256 msec temporal cadence, thereby facilitating the tracking of the radio source path from the lower to the upper atmosphere (see equation \ref{equation1}). These observations provide valuable insights into the location and temporal evolution of solar radio bursts.

\subsection{EUV and X-Ray observations }

Radio observations offer excellent temporal resolution and are often the earliest indicator of rapid coronal disturbances, making them highly valuable for identifying the onset and evolution of transient phenomena such as solar flares, CMEs, and shock waves. Despite this, Radio observations alone cannot provide a complete picture of the event because of their relatively poor spatial resolution compared to EUV and X-ray imaging observations. 

To overcome this, we complement the radio data with EUV and magnetic field data. The EUV observations were obtained from the Solar Dynamic Observatory \citep[SDO;] []{2012SoPh..275....3P}, particularly using the Atmospheric Imaging Assembly \citep[AIA;][]{Lemen2012}, which provides full-disk, multi-wavelength imaging of the atmosphere covering a wide range of plasma temperatures from the chromosphere to the corona. AIA has a pixel scale of approximately 0.6 arcsec per pixel with a temporal cadence of 12 s, enabling detailed tracking of plasma heating, loop evolution, and eruptive structures associated with the event. In addition, the Helioseismic and Magnetic Imager \citep[HMI;][]{2012SoPh..275..207S} on SDO was used to study the evolution of the photospheric magnetic field within the active region. HMI provides high-resolution line-of-sight and vector magnetic field measurements with a spatial resolution of about 1 arcsec and a temporal cadence of 45 s (vector products typically at 12-minute intervals), providing detailed measurements of changes in the magnetic topology associated with magnetic energy buildup and release. 

Further, to obtain a different perspective on the eruption, we used the EUVI maps. EUVI is part of the SECCHI instrument suite on board STEREO, which provides full-disk images of the solar corona, similar to those from SDO/AIA. The temporal and spatial resolutions are $\sim 5-10$ min and 1.6 arcsec, respectively. Although its cadence is lower than AIA, its stereoscopic viewpoint is crucial, and when combined with other perspectives, it enables reconstruction of the three-dimensional structure of the coronal loops and eruptive events. 

In the present case, enhanced brightening and a small FR eruption were observed during the flare evolution phase in both the EUVI/STEREO-A and SDO/AIA across all spectral channels (shown in the top panel of Figure \ref{Fig:figure3}). From the Earth's perspective, the event occurred near $14^\circ$ East heliographic longitude, indicating that it was observed as a disk event. To further examine the viewing geometry, we analyzed the event from STEREO-A's perspective. At the time of the event, STEREO-A was located approximately $27.7^\circ$ west of the Earth-Sun line. Using simple geometric considerations, we estimated the angular separation between the active region and the STEREO-A viewing direction. The resulting angle was approximately $42^\circ$, indicating that the eruption also appeared as a disk event from the STEREO-A viewpoint (see Figure \ref{fig:spacecraft}).

High-frequency radio and hard X-ray emissions are complementary signatures of flare-accelerated electrons. To investigate the link between the observed radio emission and high-energy electron acceleration, observations from the HELIOS/Aditya-L1 and the Spectrometer Telescope for Imaging X-rays \citep[STIX;][]{2020A&A...642A..15K} on board SolO were analyzed. Although STIX provides both imaging and spectroscopic capabilities, its automatic attenuator is inserted during intense X-class flares to mitigate detector saturation by preferentially attenuating low-energy X-ray photons, thereby affecting the low-energy spectral response. Since the present study requires reliable spectral fitting to determine the non-thermal break energy, the HELIOS observations available for this event were used for the spectroscopic analysis. As HELIOS does not provide imaging observations, STIX was used to determine the spatial distribution of the hard X-ray emission. Despite attenuation affecting its low-energy spectral response, STIX's imaging capabilities remain suitable for localized (30-50 keV) high-energy hard X-ray sources during intense flares. Furthermore, during the event, the angular separation between Solar Orbiter and the Earth line of sight was only $\sim-6.3^\circ$, indicating that both instruments observed nearly the same viewing perspective. STIX imaging observations were therefore used to identify the flare footpoint and loop--top emission regions. The footpoint sources trace the precipitation of accelerated electrons into the dense chromosphere, whereas the looptop source corresponds to hard X-ray emission from energetic electrons in the coronal flare loop \citep{2022hxga.book...96H}.

The HEL1OS instrument provides continuous, time-resolved spectroscopy of solar flare X-ray emission over the 8--150 keV energy range using compound semiconductor detectors, namely CdTe (8--70 keV) and CZT (20--150 keV). This broad spectral coverage enables simultaneous measurements of low- and high-energy X-ray photons, with spectral resolutions of approximately 1 keV at 14 keV for CdTe and 7 keV at 60 keV for CZT, allowing a clear separation of thermal and non-thermal emission components. Complementing the spectral observations, the STIX hard X-ray imaging spectrometer is designed to observe solar flares and diagnose both thermal and non-thermal electron populations via bremsstrahlung. It operates in the energy range of approximately $4$-$150$ keV, consists of multiple sub-collimators, each of which samples a specific Fourier component of the flaring X-ray source \citep{2020A&A...642A..15K}. 
In this study, we utilize the MEM\_GE \citep[][]{2020ApJ...894...46M} algorithm, which is based on a convex optimization approach and provides stable reconstructions with reduced imaging artifacts. In addition, MEM\_GE offers improved special resolution and preserves compact source structures, making it particularly suitable for accurately localizing hard X-ray sources associated with non-thermal electrons. 

\subsection{White--light observations }
To investigate whether the type II radio burst was associated with a CME, we examined the coronagraph observations from multiple instruments. Coronagraphs use an occulting disk to block the bright solar photosphere, thereby allowing observations of the faint outer corona and CME structures. Specifically, we used data from LASCO/SOHO and the SECCHI COR1 and COR2 coronagraphs on board STEREO. These coronagraphs provide observations over different heliocentric distance ranges: LASCO C2 and C3 cover approximately 2--6 and 3.7--30~$R_{\odot}$, respectively, while STEREO COR1 and COR2 cover about 1.5--4 and 2.5--15~$R_{\odot}$. The combined observations from these instruments provide complementary coverage of the corona from the low to outer regions, enabling the tracking of CME initiation and propagation over a broad range of heliocentric distances.

For this event, no clear CME signature was detected within the field of view (FOV) of any of the coronagraphs (shown in the bottom panel of Figure \ref{Fig:figure3}). However, an eruption of plasma material, together with a small FR, was observed at approximately $\sim$ 13:40 UT in multiple EUV channels in the low corona. The absence of a detectable CME makes this event particularly interesting, as type II radio bursts are typically associated with CME-driven shocks. Therefore, this study aims to identify the physical processes responsible for generating high-frequency type II radio bursts without CME association. The event timeline and corresponding observations from the different instruments are summarized in Table \ref{tab:observations}. 

\begin{figure*}[t!]
    \centering
  \includegraphics[width=0.75\textwidth]{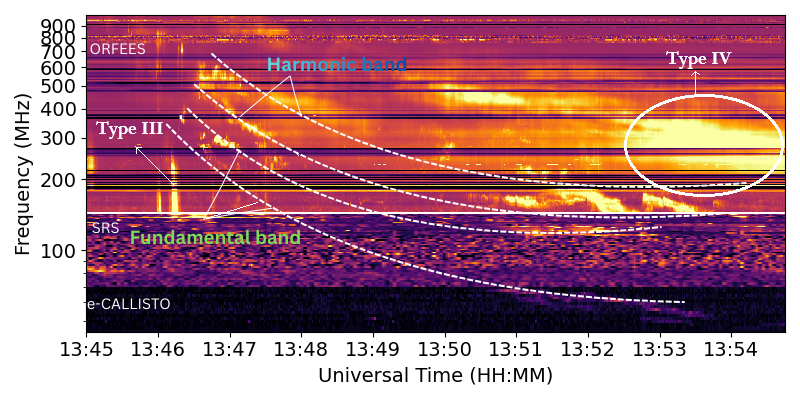}
  \includegraphics[width=0.74\textwidth]{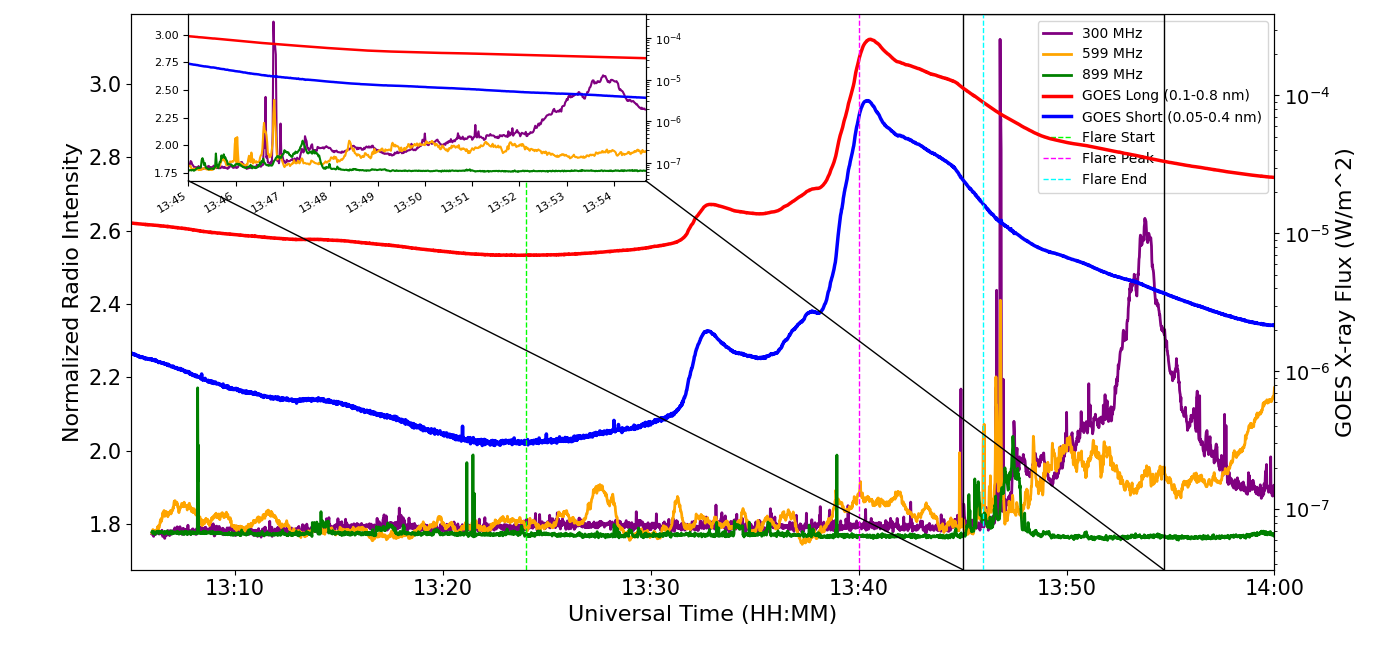}
  \vspace{-0.3cm}
  \caption{{\textbf {Top panel:}} Combined dynamic spectra of the high frequency type II radio burst observed with e-CALLISTO (GLASGOW), SRS (San Vito RSTN station, Italy), ORFEES (France) solar radio spectrographs. The dashed lines on the spectra show the fundamental (green) and harmonic (cyan) band pairs. The vertical emission lines correspond to type III radio bursts, and the enhanced emission after type II is a moving type IV radio burst. \textbf{Bottom panel:} Normalized radio intensity profile at 300 MHz (purple), $\sim 600$ MHz (orange), and $\sim 900$ MHz (green), obtained from the combined ORFEES, CALLISTO, and SRS observations during type II burst. The X-ray flux measured by the GOES-16 satellite in the 0.05--0.4 nm (orange) and 0.1--0.8 nm (blue) channels overlaid on the right axis. The lime, magenta, and cyan dashed vertical lines indicate the flare start, peak, and end times, respectively. The inset (top left panel) shows an enlarged view of the interval from 13:35:00 UT to 13:54:40 UT, highlighting the nearly simultaneous intensity peaks at 300 MHz and $\sim 600$ MHz during the high-frequency type II burst associated with the X2.3-class flare on 6 November 2024.}
  \label{Fig:figure1}
\end{figure*}

\begin{figure*}[t!]
    \centering

    \includegraphics[width=0.95\textwidth]{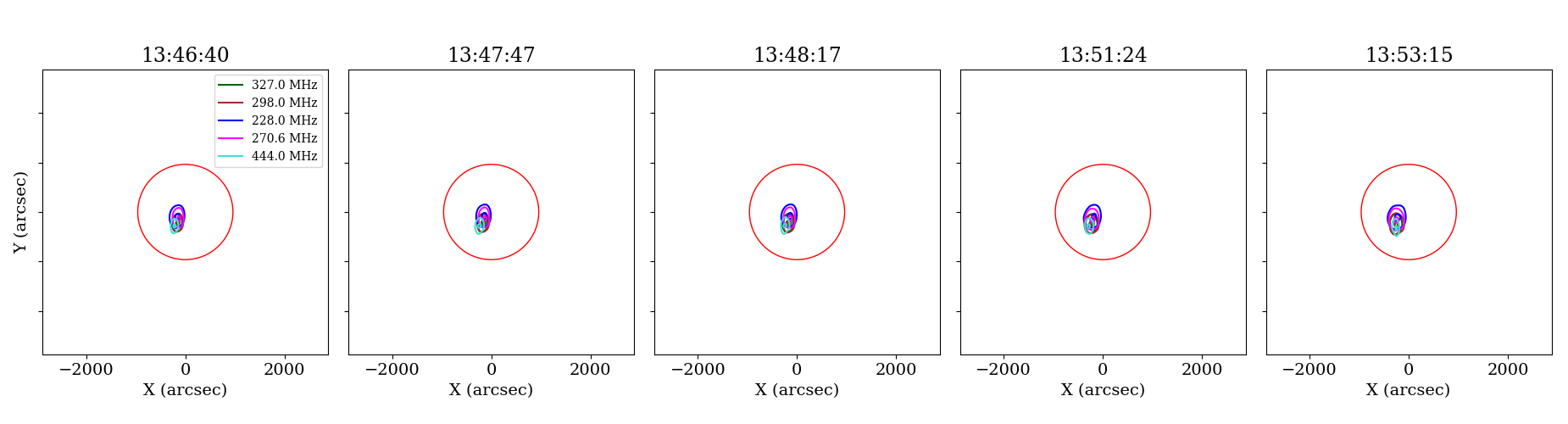}

    \vspace{0.3cm}
    \includegraphics[width=0.9\textwidth]{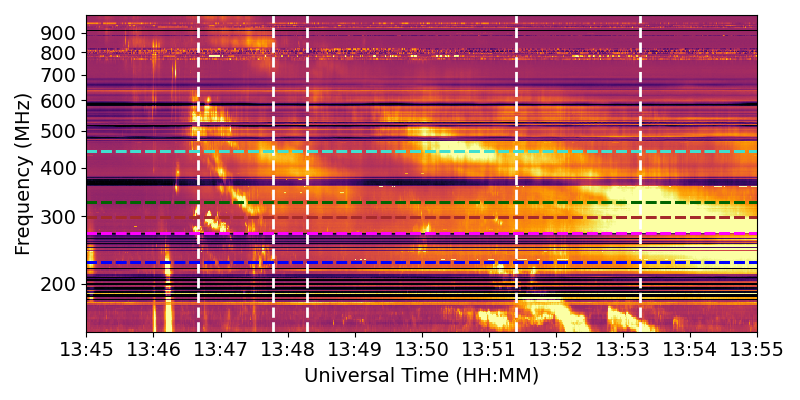}

    \caption{\textbf{Top panel:} NRH radio source contours at 50 \% and 90 \% intensity levels for the selected times and frequencies. The red circle denotes the projected solar disk. The radio sources remain nearly co-spatial throughout the event and are concentrated above the flare-producing active region. \textbf{Bottom panel:} Dynamic radio spectrum of observed bursts by ORFEES. The horizontal dashed line indicates the chosen frequencies used for NRH imaging: 327.0 MHz (green), 298.0 MHz (brown), 228.0 MHz (blue), 270.6 MHz (magenta), and 444.0 MHz (cyan). The vertical dashed line marks the corresponding imaging times (13:46:40, 13:47:47, 13:48:17, 13:51:24, and 13:53:15 UT) shown in the top panel. }
    \label{Fig:figure2}
\end{figure*}

{\color{red}

\begin{table*}[t]
\centering
\caption{Summary of the multiwavelength observations associated with the 6 November 2024 event.}
\label{tab:observations}

\begin{tabular}{lccc}
\hline
Instrument & Start Time (UT) & End Time (UT) & Observation \\
\hline
GOES               & 13:24 & 13:46 & X2.3 flare \\
Radio              & 13:46 & 13:56 & High-frequency Type II burst \\
SDO/AIA            & 13:39 & ---   & EUV brightenings and small flux rope \\
STEREO-A/EUVI      & 13:40 & ---   & Coronal loop \\
STEREO-A/COR1      & ---   & ---   & No white-light CME detected \\
STEREO-A/COR2      & ---   & ---   & No CME detected \\
SOHO/LASCO C2      & ---   & ---   & No CME detected \\
SOHO/LASCO C3      & ---   & ---   & No CME detected \\
\hline
\end{tabular}

\end{table*}

}

{\color{red}

\begin{figure}[t!]
\centering
\includegraphics[width=0.70\linewidth]{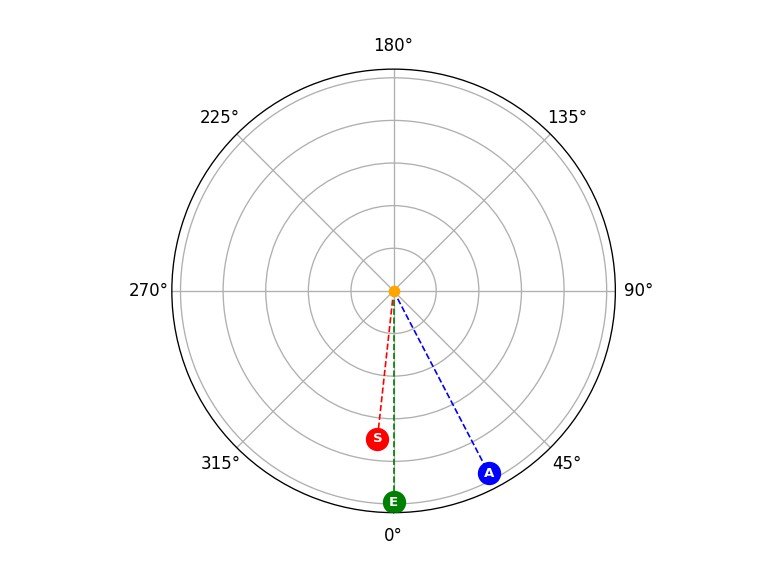}
\caption{Spacecraft configuration on 6 November 2024 in heliographic coordinates. The positions of Earth (green), STEREO-A (blue), and Solar Orbiter (red) are shown in the ecliptic plane. The dashed lines indicate the Sun--spacecraft directions \citep{2023FrASS...958810G}.}
\label{fig:spacecraft}
\end{figure}

}

\section{analysis and results} \label{sec:analysis}

\subsection{Spectral and Imaging analysis of the high--frequency type II radio burst}
Figure \ref{Fig:figure1} depicts the evolution of a high-frequency type II radio burst. The upper panel shows the dynamic spectrum, where the type II starts at $\sim 750$ MHz around 13:46 UT and drifts to $\sim 45$ MHz by 13:56 UT. The burst exhibits slowly drifting emission lanes corresponding to the fundamental ($F_f$) and harmonic ($F_h$) bands. The harmonic component is identified by the characteristic frequency ratio of approximately 1:2 ($ F_h$ = 2$F_f$). The burst occurs during the decay phase of the flare and exhibits a frequency drift rate of $\sim 0.5$ MHz/s, consistent with the outward propagation of a coronal shock through regions of decreasing plasma density. The lower panel displays the corresponding time profiles extracted from the combined spectra of ORFEES, CALLISTO, and SRS observations at 300 MHz (purple), $\sim600$ MHz (orange), and $\sim900$ MHz (green). The radio intensity profiles are overlaid with the GOES-16 X-ray light curves in the 0.1--0.8 nm (red) and 0.05--0.4 nm (blue) channels to compare the temporal evolution of the radio emission with that of the associated flare. The flare start, peak, and end times are indicated by the lime, magenta, and cyan dashed vertical lines, respectively. While the GOES X-ray flux shows the characteristic rapid rise followed by a gradual decay, the radio emission exhibits frequency-dependent variability. The 300 MHz profile displays several impulsive enhancements followed by a pronounced increase in intensity after the flare maximum, whereas the $\sim600$ MHz profile shows comparatively weaker fluctuations, and the $\sim900$ MHZ profile remains nearly constant except for a few short-duration bursts. A zoomed-in inset of the interval between 13:35:00 UT and 13:54:40 UT, shown in the upper-left corner, provides a clearer view of the radio source's temporal evolution. The enlarged profile shows that the intensity enhancement at 300 MHz and $\sim600$ MHz occurs nearly simultaneously. This close temporal correspondence, together with their approximately 1:2 frequency ratio, supports their identification as the fundamental and harmonic components of the type II radio burst. Further, to investigate a possible third harmonic, we performed a crossed-cut at 900 MHz; however, the splitting peak was not visible, indicating it was not a third-harmonic peak.

In addition to these datasets, we utilized the Nançay Radioheliograph (NRH) for radio imaging observations. The NRH produces two-dimensional solar radio images across ten different frequency channels: 150.9, 173.2, 228.0, 270.6, 298.7, 327.0, 382.2, 408.0, 432, and 444.0 MHz. Radio imaging is used to track the evolution of the radio-emission sources associated with type II bursts. We can monitor the radio source centroid from the pre-flare phase to the decay phase. The centroid size appears smaller at higher frequency (indicating lower solar atmospheric heights), while its size appears larger at lower frequency (indicating higher solar atmospheric heights). This allows us to identify the spatio-temporal behavior of the radio source, whereby higher-frequency radio emission originates from regions of higher density and lower-frequency emission from regions of lower density. 

The upper panel of Figure \ref{Fig:figure2} shows the NRH radio source contours at the 50\% and 90 \%, intensity levels for five selected frequencies: 228.0 MHz (blue), 270.6 MHz (magenta), 298.0 MHz (brown), 327.0 MHz (green) and 444.0 MHz (cyan), overlaid at five different times 13:46:40, 13:47:47, 13:48:17, 13:51:24, and 13:53:15 UT, respectively. These frequencies and times were chosen from the zoomed--in time--frequency spectrum shown in the bottom panel of the Figure \ref{Fig:figure2}, where the horizontal dashed line indicates the selected frequencies and the vertical dashed lines mark the corresponding imaging times. The NRH images reveal that the radio source remains nearly stationary in the plane of the sky throughout the bursts. Since the event originated near the disk center, a predominantly radial eruption would be expected to propagate largely along the line of sight. Consequently, any radial outward motion is strongly affected by projection, producing only a small displacement of the radio source in the plane of the sky. Furthermore, the spatial resolution of the NRH observations may not be sufficient to resolve small height-dependent shifts in the radio source location. Therefore, complementary multiwavelength observations are required to constrain the eruption geometry and identify the physical driver of the radio emission.

\subsection{Tracing coronal disturbances: EUV wave signatures without a white--light CME}

Figure \ref{Fig:figure3} shows snapshots of the compact bright front associated with the rising magnetic FR, as observed by SDO/AIA and STEREO-A/EUVI. The left panels present observations from SDO/AIA 171 Å images, while the right panels show STEREO-A/EUVI 195 Å images. Two different times are displayed to show the evolution and expansion of the rising FR. The expanding wavefronts and associated coronal disturbances, interpreted as manifestations of the EUV wave, are indicated by magenta arrows that trace their propagation from the eruption site.

Although plasma is seen erupting from the active region in EUV observations, no corresponding large-scale white-light CME was detected in STEREO/COR1, COR2, or SOHO/LASCO-C2/C3 observations (bottom panel of the Figure \ref{Fig:figure3}). The EUV eruption and wavefront marked by magenta arrows were clearly visible in all channels of the SDO-AIA and STEREO-EUVI instruments (top panel of the Figure \ref{Fig:figure3}). Only a few studies have reported shock formation in the absence of a clearly observed CME, with the shock associated with an outward-propagating EUV wave at 400-600 km/s, as well as with the jet, high-pressure flare blast waves, and failed eruptions \citep{magdalenic2012flare, SU2015,2023arXiv230511545M, 2025JApA...46...90K}. The EUV wave represents a large-scale propagating disturbance in the solar corona. Such EUV waves are widely interpreted as fast-mode MHD waves or shocks, while metric type II radio bursts are generally treated as signatures of coronal shock waves propagating through the solar atmosphere \citep{2015LRSP...12....3W}. In the present event, the close temporal association between the propagating EUV disturbance and the onset of the type II burst suggests that the outward-propagating EUV wave evolved into a coronal shock. Such a shock can accelerate electrons through plasma emission processes, producing the observed high-frequency type II radio burst.

\begin{figure}[t!]

\includegraphics[width=1\linewidth]{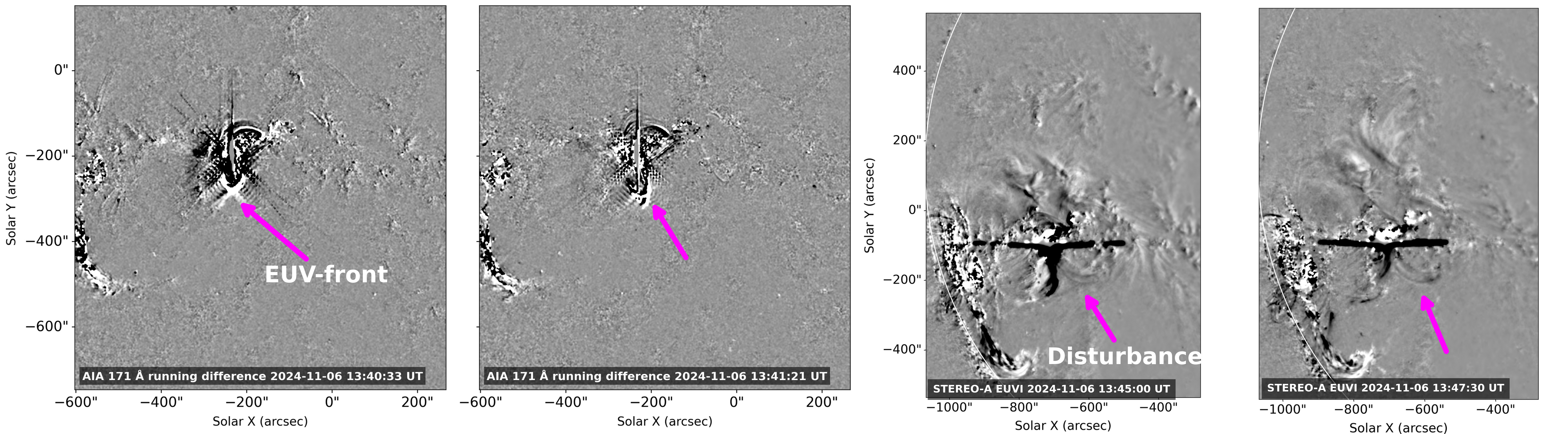}

\vspace{0.2cm}

\includegraphics[width=1\linewidth]{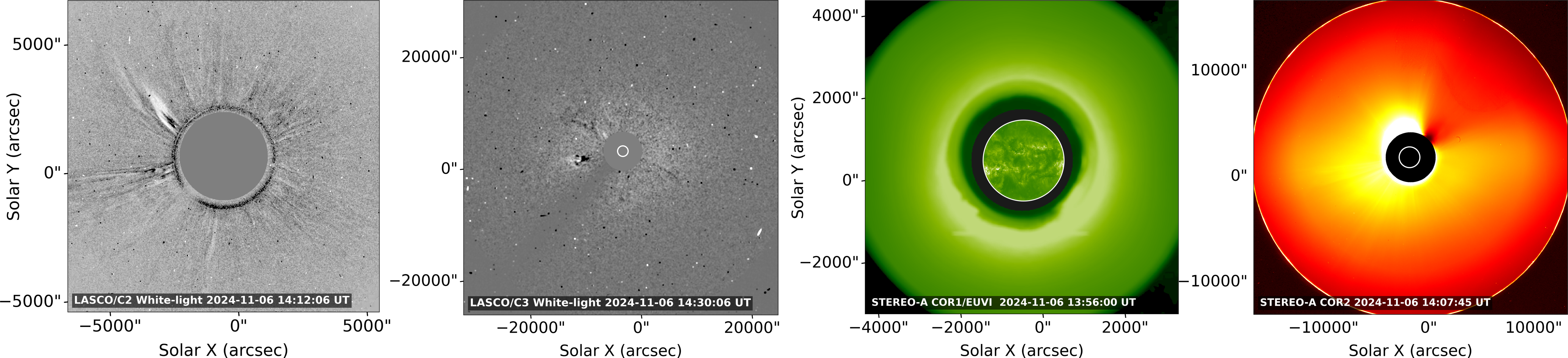}
\caption{\textbf{Top panel:} Running-difference images from SDO/AIA 171 \AA\ (right; 13:40:33 UT and 13:41:21 UT) and STEREO-A/EUVI 195 \AA\, 171 \AA\ (left; 13:40:00 UT and 13:47:30 UT) showing the evolution of the erupting FR and associated propagating EUV wavefront and coronal disturbance. The propagation of the EUV wavefront is indicated by magenta arrows. \textbf{Bottom panel:} White--light coronagraph images from SOHO/LASCO C2, and SOHO/LASCO C3 (left; 14:12:06 UT and 14:30:06 UT), STEREO-A/COR1 and STEREO-A/COR2 (right; 13:56:00 UT and 14:07:45 UT). Despite the presence of a large-scale EUV wavefront and disturbance, no CME signature is detected in white-light coronagraph observations. 
}
\label{Fig:figure3}

\end{figure}

\subsection{Estimation of the average shock speed using EUV--AIA/SDO and from radio dynamic spectra}

The true heliocentric height of the erupting structure was determined using stereoscopic triangulation of simultaneous observations from SDO/AIA 211\,\AA\ and STEREO-A/EUVI. The outermost apex of the expanding feature was identified in both viewpoints, and its three-dimensional position was reconstructed using the triangulation and tie-pointing routine of \citep{2023SoPh..298...36N} in the Heliocentric Earth Equatorial (HEEQ) coordinate system. The erupting structure was identified in only two EUVI frames, one in the 195\,\AA\ channel and one in the 171\,\AA\ channel, and these were used for stereoscopic triangulation with the nearest--in--time AIA 211\,\AA\ observations. The first stereoscopic pair consisted of the AIA 211\,\AA\ frame at 13:39:57~UT and the EUVI 195\,\AA\ frame at 13:40:00~UT, yielding a heliocentric height of the apex of $1.029\, R_{\odot}$. The second stereoscopic pair consisted of the AIA 211\,\AA\ frame at 13:40:37~UT and the EUVI 171\,\AA\ frame at 13:40:30~UT, for which the heliocentric height of the apex increased to $1.055\, R_{\odot}$. Using the average observation times of the two stereoscopic measurements, separated by 34~s, we derive the 3D velocity of $\sim 536 \pm 57~\mathrm{km\,s^{-1}}$. The uncertainty was estimated from fifteen independent tie-point selections.

\begin{figure*}[ht!]
  \includegraphics[width=0.5\textwidth]{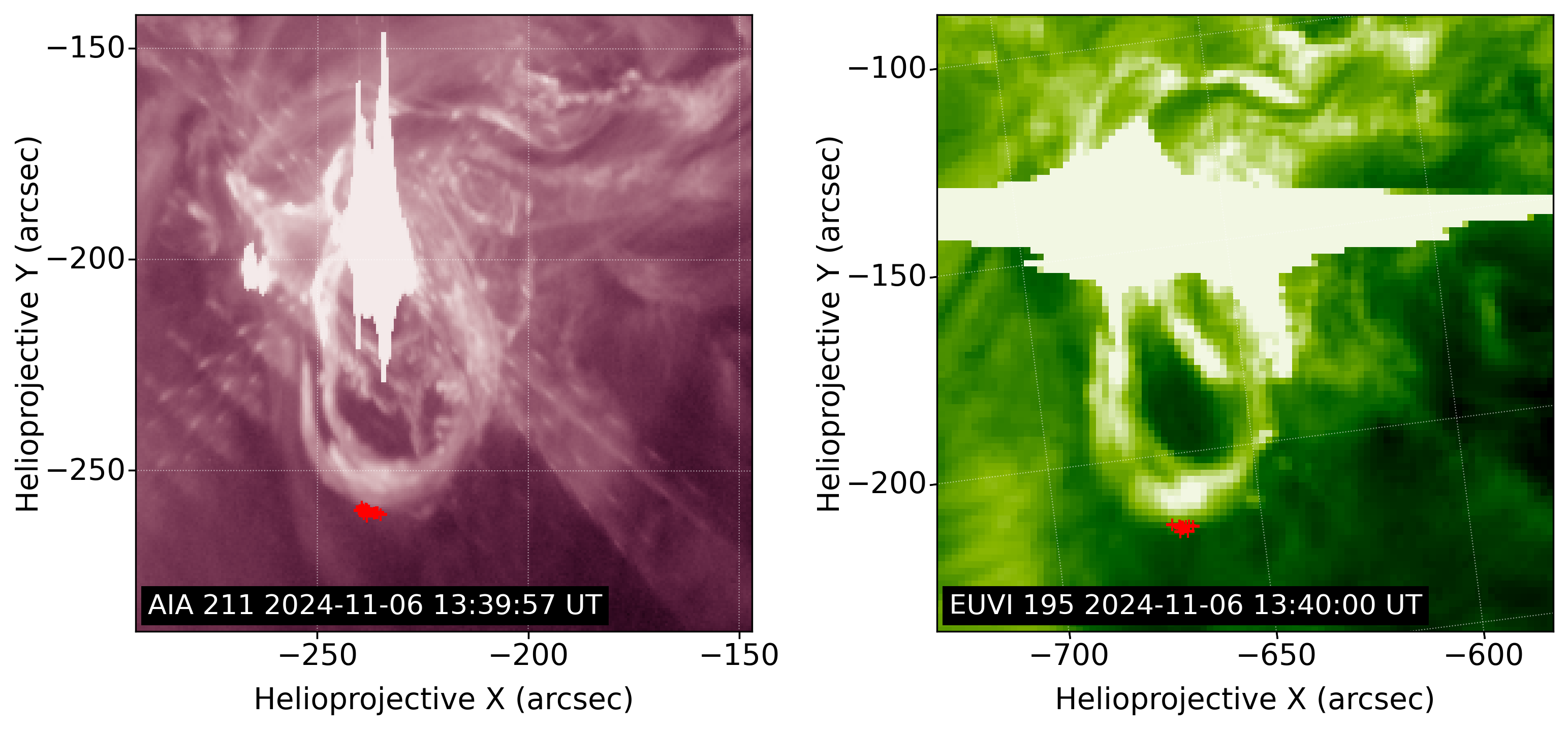}
  \includegraphics[width=0.5\textwidth]{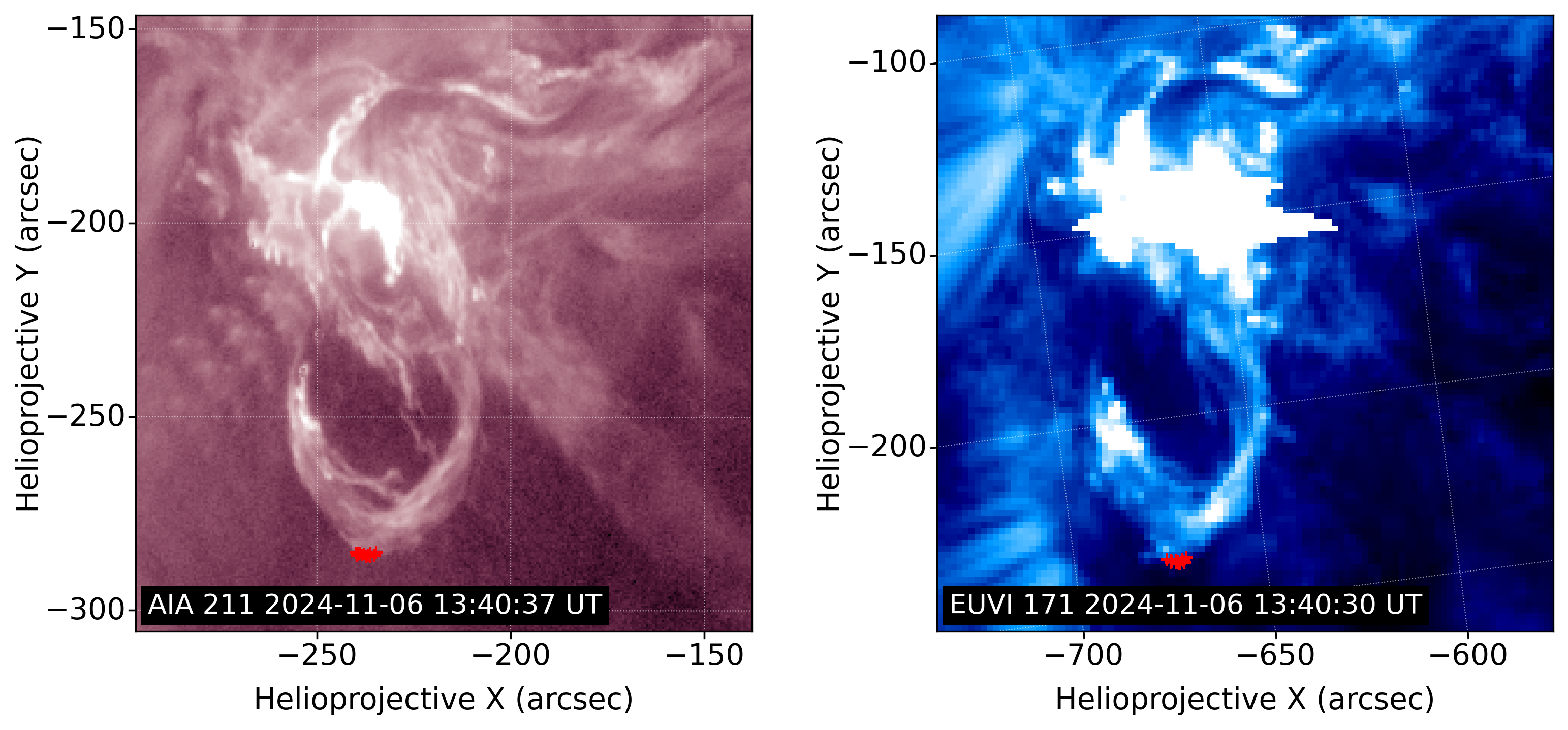}
  \vspace{-0.3cm}
  \caption{{Zoomed stereoscopic snapshots showing the evolution and expansion of the erupting FR. The left panels correspond to SDO/AIA 211 Å and STEREO-A/EUVI 195 Å observations at 13:39:57--13:40:00 UT, while the right panels correspond to SDO/AIA 211 Å and STEREO--A/EUVI 171 Å observations at 13:40:37--13:40:30 UT. The average time difference between the stereoscopic snapshots is 34 s, yielding an estimated radial velocity of $\sim 536 \pm 57~\mathrm{km\,s^{-1}}$. The red (+) markers indicate the fifteen independent tie-point selections used to quantify the uncertainty in the triangulated position.} }
  \label{Fig:figure4}
\end{figure*}

The frequency drift observed in the dynamic spectra indicates the speed of the shock wave propagating into the solar corona. To calculate the shock speed, it is essential to understand the distribution of the corona's density. However, numerous density models have been proposed by various researchers, each accounting for different physical scenarios. Among the available coronal density models, the Newkirk model is among the most widely used in previous studies \citep{2009SoPh..259..227G, Ma2011}. It provides a reasonable representation of active-region coronal densities and has been extensively adopted in studies of type II radio bursts.  

Since the electron density above active regions is generally higher than that of the quiet-Sun corona, the Newkirk model is commonly modified by applying a constant enhancement factor, typically taken as 2. The emission frequency of a type II radio burst corresponds to the local electron plasma frequency or its second harmonic, produced by electrons accelerated at a propagating coronal shock. Using the adopted density model, the observed frequency drift can then be converted into a height--time profile, from which the shock speed is determined.

\begin{equation}
    f_{p} \approx 9 \times \sqrt{n_{e}} \quad \text{(in kHz)}
    \label{equation1}
\end{equation}

\begin{equation}
n_{e}(r) = E \times 4.2 \times 10^{4} \, 10^{\frac{4.32}{r}}
\label{equation2}
\end{equation}

where \(n_e\) is the electron density, \(E\) is the  
enhancement factor, and \(r\) is the heliocentric distance 
in units of \(R_{\odot}\).

\vspace{0.4cm}
According to the Newkirk density model, there is a relationship between the burst frequency and the density in the shock region, as well as the corresponding height \citep{newkirk1961solar}. This allows us to estimate the shock speed from the frequency drift rate. We manually selected 20 points that mark the start and end times of type II bursts in both the fundamental and harmonic bands. It is observed that the fundamental band becomes significantly fainter at lower frequencies. The onset of the type II burst approximately 360 s after the first appearance of the EUV disturbance is consistent with the formation of a coronal shock during the evolution of the FR. The EUV wave propagates through the corona, driving localized coronal disturbances, and a type II radio signature is subsequently observed in the dynamic spectra. The frequency--time profile was converted into a height--time profile using equations  \ref{equation1} and \ref{equation2}.  
The results indicate that the average shock speed was approximately 540 ± 39 km/s (see Figure \ref{Fig:figure5}). Noted that this speed is consistent with the speed of the erupting FR estimated from AIA/SDO and EUVI/STEREO images using the triangulation method.

\begin{figure}[t!]
    \centering
    \includegraphics[width=0.5\linewidth]{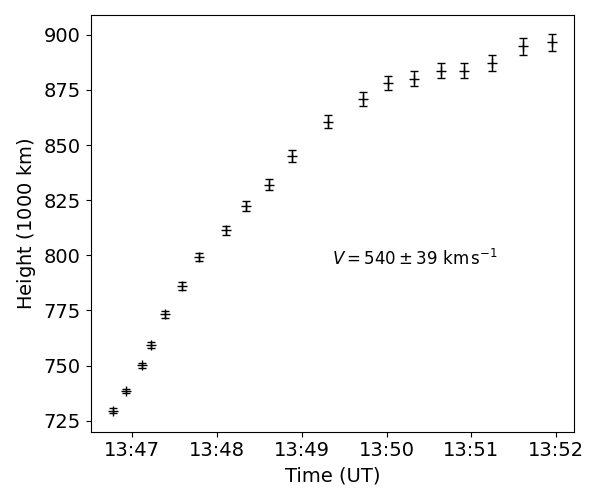} 
    \caption{ Height-- time evolution of type II radio burst source derived from the dynamic spectrum during 13:46:40–13:52:07 UT. The source heights were estimated by converting the observed emission frequencies to coronal heights using the Newkirk density model. The vertical error bars represent the $1\sigma$ uncertainties in the estimated height. A linear fit to the height--time measurements yields a shock speed $\sim 540 \pm 39~\mathrm{km\,s^{-1}}$. }
    \label{Fig:figure5}
\end{figure}

\subsection{Non-thermal electron energy estimation using HEL1OS and X-ray counterparts} \label{sec: X-ray}

To investigate the plasma heating and electron acceleration associated with the flare, we analyze X-ray spectroscopy obtained by HEL1OS onboard Aditya-L1 together with hard X-ray imaging from STIX onboard Solar Orbiter. X-ray spectroscopy provides quantitative information on the thermal and non-thermal properties of the emitting plasma, while hard X-ray imaging identifies the locations of energetic electron interactions during the flare. These complementary observations are used to characterize the energetic particle population associated with the eruption and to place the radio observations in a broader multiwavelength context. The HEL1OS spectrum obtained during 13:38:00--13:38:40 UT, corresponding to the rise phase of the flare, is presented in the left panel of the Figure \ref{Fig:figure6}. Although HEL1OS covers the 8–150 keV energy range, the spectral fitting was performed above 10 keV because the 8–10 keV interval is not yet sufficiently modeled. The fit was extended to the highest energy with an adequate signal-to-noise ratio. The spectrum was fitted in XSPEC\footnote{\url{https://heasarc.gsfc.nasa.gov/docs/software/xspec/}} using a combined chisoth\footnote{\url{https://github.com/xastprl/chspec}} and broken power law (bknpower) models \citep{1996ASPC..101...17A}. The chisoth component, based on the CHIANTI atomic database, describes the continuum and line emission from an optically thin, ionized plasma, with the plasma temperature, emission measure, and elemental abundances as its free model parameters \citep{2021ApJ...920....4M}. The bknpower component represents the non-thermal bremsstrahlung produced by flare-accelerated electrons and is characterized by the photon spectral indices, normalization, and break energy that describes the transition between the two power-law regimes \citep{ 2008AdSpR..42..828K, 2011A&A...529A.109H,2021ApJ...920...41M}.

The left panel of Figure \ref{Fig:figure6} shows the X-ray spectral fitting plot, where the orange and green data points correspond to the CdTe and CZT detector spectra, respectively. The blue dotted curve represents the thermal chisoth component, the magenta dashed curve shows the non-thermal bknpower component, and the solid red curve denotes the combined best-fit model. The bottom panel displays the normalized residuals ($\Delta\chi$), which are randomly distributed about zero, indicating that the adopted model provides an excellent representation of the observed spectrum with a reduced chi--square of $\chi_r^2 \approx 0.9$. The fitted spectrum clearly demonstrates the coexistence of thermal and non-thermal emission. The low-energy emission is well reproduced by an isothermal plasma at $T \approx 23.31$ MK, indicating substantial plasma heating during the flare rise phase. At higher energies, the spectrum is dominated by a non-thermal component with a photon spectral index of $\gamma = 4.18$, providing strong evidence for efficient electron acceleration. The fitted break energy, $E_b = 17.36$ keV, is a parameter of the broken power-law component. Since XSPEC does not provide a model with an explicit low-energy cutoff, the fitted break energy serves as a close approximation to the cutoff energy of the accelerated electron distribution, representing the minimum energy above which electrons contribute significantly to the observed hard X-ray emission.

\begin{figure*}[h!]
    \centering

    \begin{minipage}{0.48\textwidth}
        \centering
        \includegraphics[width=\linewidth]{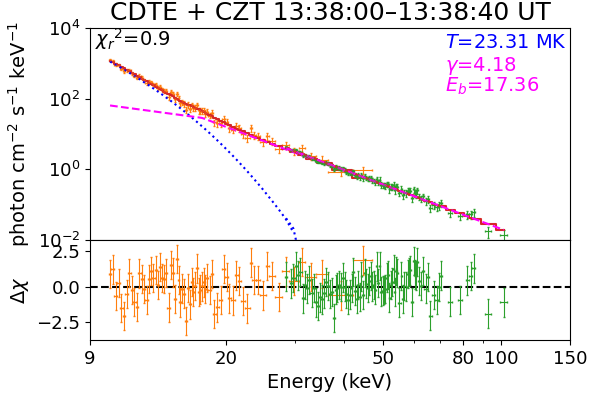}
    \end{minipage}
    \hfill
    \begin{minipage}{0.48\textwidth}
        \centering
        \includegraphics[width=\linewidth]{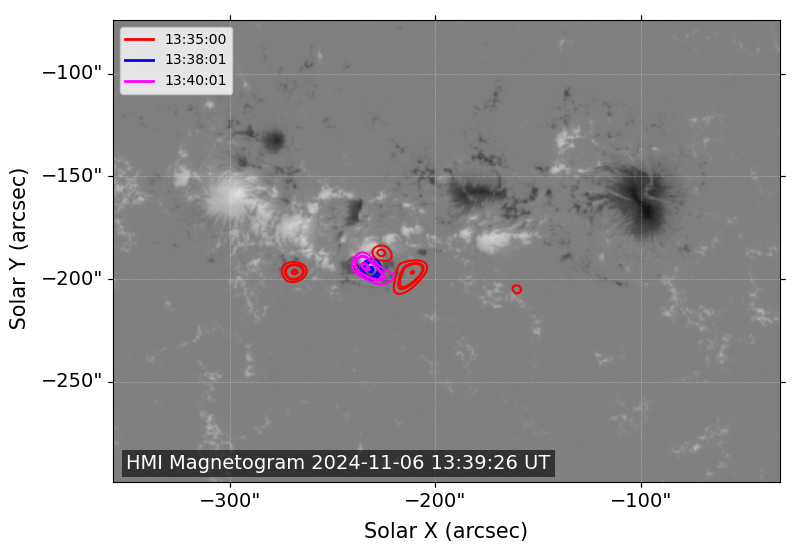}
    \end{minipage}

    \vspace{1mm}

    \caption{
\textbf{Left panel:} HELIOS onboard Aditya-L1 X-ray photon spectra from 13:38:00--13:38:40 UT on 6 November 2024, fitted with a combined thermal and non-thermal model. The upper panel shows the observed hard X-ray spectrum together with the best-fit model, while the lower panel shows normalized residuals. The blue dashed line represents the thermal (chisoth) components, the magenta dashed line represents the non-thermal (bknpower) component, and the solid dark red curve represents their combined fit. The derived parameters are $T = 23.31~\mathrm{MK}$, $\gamma = 4.18$, and $E_b = 17.36~\mathrm{keV}$.
\textbf{Right panel:} STIX onboard Solar Orbiter hard X-ray contours overlaid on SDO/HMI line-of-sight magnetogram during the flare on 6 November 2024. Contours are shown at 50\%, 70\%, and 90\% of the peak intensity for 13:35:00 UT (red), 13:38:01 UT (blue), and 13:40:01 UT (magenta), within a 30--50 keV range.
}

   \label{Fig:figure6}
\end{figure*}
The time-resolved spectral analysis indicates that the low-energy cutoff of the non-thermal electron distribution is approximately 17.36 keV, implying that the non-thermal contribution to the X-ray emission becomes significant above this energy. In addition, the attenuator configuration during the flare substantially improves the imaging performance of STIX at energies above $\sim 25$ keV. Consequently, STIX images were reconstructed in the 30-50 keV energy band, where the X-ray emission is expected to be predominantly nonthermal. This energy range is therefore well-suited for imaging the sites of energetic acceleration and precipitation. The reconstructed STIX images primarily trace the locations of non-thermal electron interactions, typically observed at the flare footpoints and the looptop regions during the rise, peak, and decay phases of the event, where significant energy deposition occurs.  The right panel of Figure \ref{Fig:figure6} shows the MEM\_GE reconstructed STIX 30--50 keV contours overlaid on the SDO/HMI line--of--sight magnetogram at 13:35:00 UT, 13:38:01 UT, and 13:40:01 UT. The red contours mark the compact footpoint hard X-ray sources, whereas the blue and magenta represent the loop--top emissions. The presence of both footpoint and loop-top hard X-ray sources is consistent with non-thermal electrons accelerated in the coronal flare region and subsequently propagating along magnetic loops, producing hard X-ray emission through non-thermal bremsstrahlung \citep{Krucker2010,Fleishman2011,Krucker2014,Fleishman2022}. 

Thus, the combination of X-ray diagnostics, which trace downward-moving electrons, and radio observations, which trace upward-moving electrons, provides a consistent and complete picture of the electron acceleration and transport processes during the event. This multi-instrument approach is crucial for understanding the origin and propagation of the shock responsible for the type II radio burst, and for establishing a physical connection between the EUV wave, electron acceleration, and radio emission in the absence of a CME signature. In the following, we discuss the implications of these results for the origin of CME-driven type II bursts and the role of small-scale eruptive structures in driving coronal shocks.

\section{Discussion} \label{sec:discussion}
The presence of both fundamental and harmonic emission bands confirms that a shock was generated in the low corona during the event. The type II burst was observed from approximately 13:46 UT to 13:56 UT over a broad frequency range. Although the fundamental lanes are not clearly distinguishable in the dynamic spectra, their identification is supported by the simultaneous intensity enhancements observed in the time profiles near the fundamental ($\sim300$ MHz) and harmonic ($\sim600$ MHz) frequencies, including split band structures. Furthermore, manual tracing of emission lanes shows that the frequency ratio $f_H$/$f_F$ remains close to 2, supporting their interpretation as the fundamental and harmonic components of the type II emission. Similar unusually high-frequency type II radio bursts have been reported by \citet{cho2013high, magdalenic2012flare}, indicating that coronal shock can be generated at relatively low coronal heights where the ambient plasma density is sufficiently high.  

In addition, the NRH radio sources observed at 228 MHz and 444 MHz are spatially associated with the same observed shock region and correspond to frequencies that satisfy the expected F-H relationship, providing independent support for their identification as the F-H pair. Both the fundamental and harmonic lanes also exhibit clear band splitting. Recent studies have suggested that such band splitting may arise from spatial variations in the ambient plasma conditions and shock geometry, including density inhomogeneities across the shock, local variations in the Alfvén speed, and localized electron acceleration at different regions of a structured shock front \citep{Morosan2025, Zucca2025}.  In particular, the shock surface may not be smooth, but instead contain rippled, wrinkled, or corrugated structures, where quasi-perpendicular regions of the shock $\theta_{BN} \sim 90^\circ$ can produce enhanced radio emission intensity \citep{Kai1969} .

Using the observed split bands, we estimate relative band split widths of $BDW_F \sim 0.24$ and $BDW_H \sim 0.26$ for fundamental and harmonic lanes, respectively. The corresponding density compression ratios were found to be $\sim 1.55$ and $\sim 1.6$, while the derived Alfvén Mach numbers are $M_A \sim 1.44$ for fundamental and $M_A \sim 1.4$ for the harmonic component. These relatively small compression ratios and Mach numbers indicate a weak-to-moderate super-Alfvénic shock, implying that only modest plasma compression was required to accelerate electrons and produce the observed high-frequency type II radio emission. Unlike the CME-driven events analysis by \citet{Bhandari2025}, where type II bursts were associated with super-critical shock ($3.8 \le M_A \le 7.7$), the present event exhibits a considerably lower Alfvén Mach number. This difference suggests that, under favorable low-coronal plasma conditions, relatively moderate Alfvénic shocks may be capable of producing high-frequency type II radio emission. Using the radio-derived shock speed, we further estimate an Alfvén speed of approximately {\boldmath$\sim361\,\mathrm{km\,s^{-1}}$}, which is substantially lower than the Alfvén speeds estimated to range from {\boldmath$2\times10^{3}$} to {\boldmath$>8\times10^{3}\,\mathrm{km\,s^{-1}}$} in the flaring volume by \citet{Kaltman2026}. Since the formation of a coronal shock depends on the local Alfvén speed rather than solely on the eruption speed, such relatively low Alfvén speeds provide favorable conditions for even moderately fast eruptive structures to become super-Alfvénic and generate type II radio bursts \citep{mann2003formation,2024RSPSA.48030950Z}.   

Moreover, another drifting emission lane near $\sim 900\,\mathrm{MHz}$ was initially considered as a possible third harmonic component. However, the absence of clear split-band structures and corresponding intensity peaks in the time profile suggests that it is unlikely to represent the third harmonic. \citet{2024A&A...690A.382K} reported that moving type IV radio bursts are closely associated with erupting magnetic FR structure. In the present event, a broad continuum-like drifting structure appears after approximately 13:52 UT at higher frequencies, following the type II burst. This continuum resembles a possible moving type IV radio burst and may likewise be associated with the continued evolution of the erupting FR. However, a comprehensive investigation of this continuum emission is left for future work, as the present study focuses on the type II radio burst and associated shock properties. 

Previous studies have demonstrated that type II radio bursts can be generated in the absence of large-scale CME. \citet{magdalenic2012flare} interpreted the high-frequency type II bursts as a flare blast wave generated by the impulsive flare energy release, concluding that no evidence for piston-driven eruptive structure was present during the event. Although \citet{Kumar2022} reported an erupting FR and evidence for breakout reconnection, they found no convincing evidence that the FR acted as the piston driving the shock, owing to the absence of strong lateral expansion and a clearly identifiable piston-driven shock. Consequently, they favored a flare-blast-wave interpretation of the origin of the type II burst, while \citet{SU2015} proposed that the rapid expansion of magnetic loops served as the piston driving the shock. Also, \citet{Morosan2023b} interpreted a CME-less type II burst as being generated by a freely propagating EUV pressure wave that steepened into a fast-mode shock upon entering a region of relatively low Alfvén speed, demonstrating that a large-scale CME is not a necessary condition for shock formation. More recently \citet{2025JApA...46...90K} suggest that a compact low-coronal eruptive structure can generate a type II radio burst without a detectable white-light CME, highlighting that localized eruptions may serve as efficient drivers in the solar corona. Consistent with these scenarios, EUV observations from SDO/AIA and STEREO-A reveal the expansion and eruption of a small FR during the flare rise phase. Despite the presence of a strong flare and an associated EUV disturbance, no corresponding detectable large-scale CME, such as coronal dimming or a white-light counterpart, was observed. Nevertheless, a FR generated disturbance signature is visible near the flaring region, suggesting that the disturbance responsible for the type II burst originated in the low corona.

Using the stereoscopic triangulation from two viewpoints, we estimated the 3D speed of the erupting FR  $\sim 536 \pm 57~\mathrm{km\,s^{-1}}$, which is comparable to the average shock speed of $\sim 540 \pm 39~\mathrm{km\,s^{-1}}$  derived from the radio observations. The close agreement between the independently derived FR speed and the radio-derived shock speed is consistent with the interpretation that the expanding FR generated the coronal shock. As FR expanded through the surrounding corona, it generated a compressive disturbance that evolved into a shock. Unlike the flare blast-wave scenarios proposed by \citet{magdalenic2012flare, Kumar2022} and the loop-driven shock scenario proposed by \citet{SU2015}, the present observations indicate that the expanding FR is the most likely driver of the shock. This interpretation is consistent with the scenario proposed by \citet{2025JApA...46...90K}, while the stereoscopic observations provide direct kinematic evidence linking the expanding FR to the radio-derived shock. 

To further examine the energy release and electron acceleration associated with this event, we combined the HEL1OS X-ray spectral analysis with STIX hard X-ray imaging (Section \ref{sec: X-ray}). The coexistence of hot thermal plasma and a pronounced non-thermal component demonstrates that magnetic energy is distributed between plasma heating and electron acceleration during the impulsive phase of the flare. The inferred break energy of $\sim17$ keV suggests that electrons above this energy contributed significantly to the observed hard X-ray emission via non-thermal bremsstrahlung. The STIX 30--50 keV images further reveal both looptop and footpoint hard X-ray sources, consistent with electron acceleration in the corona followed by transport along magnetic field lines toward the chromosphere.

Therefore, the temporal and spatial association between the hard X-ray sources, the EUV disturbance, and the subsequent type II burst supports a shock driven by the erupting compact FR rather than by a large-scale CME. In the absence of an accompanying large-scale CME, these observations strongly favor a flare-driven origin of the shock responsible for the type II radio burst. While the X-ray observations constrain the sites of energy release and electron acceleration, they do not directly reveal the magnetic configuration for guiding the accelerated electrons. To investigate this aspect, we performed an NLFFF extrapolation to examine the magnetic connectivity of the source region and its role in the observed radio and X-ray emissions.

\begin{figure}[t!]
    \centering
    \includegraphics[width=\linewidth]{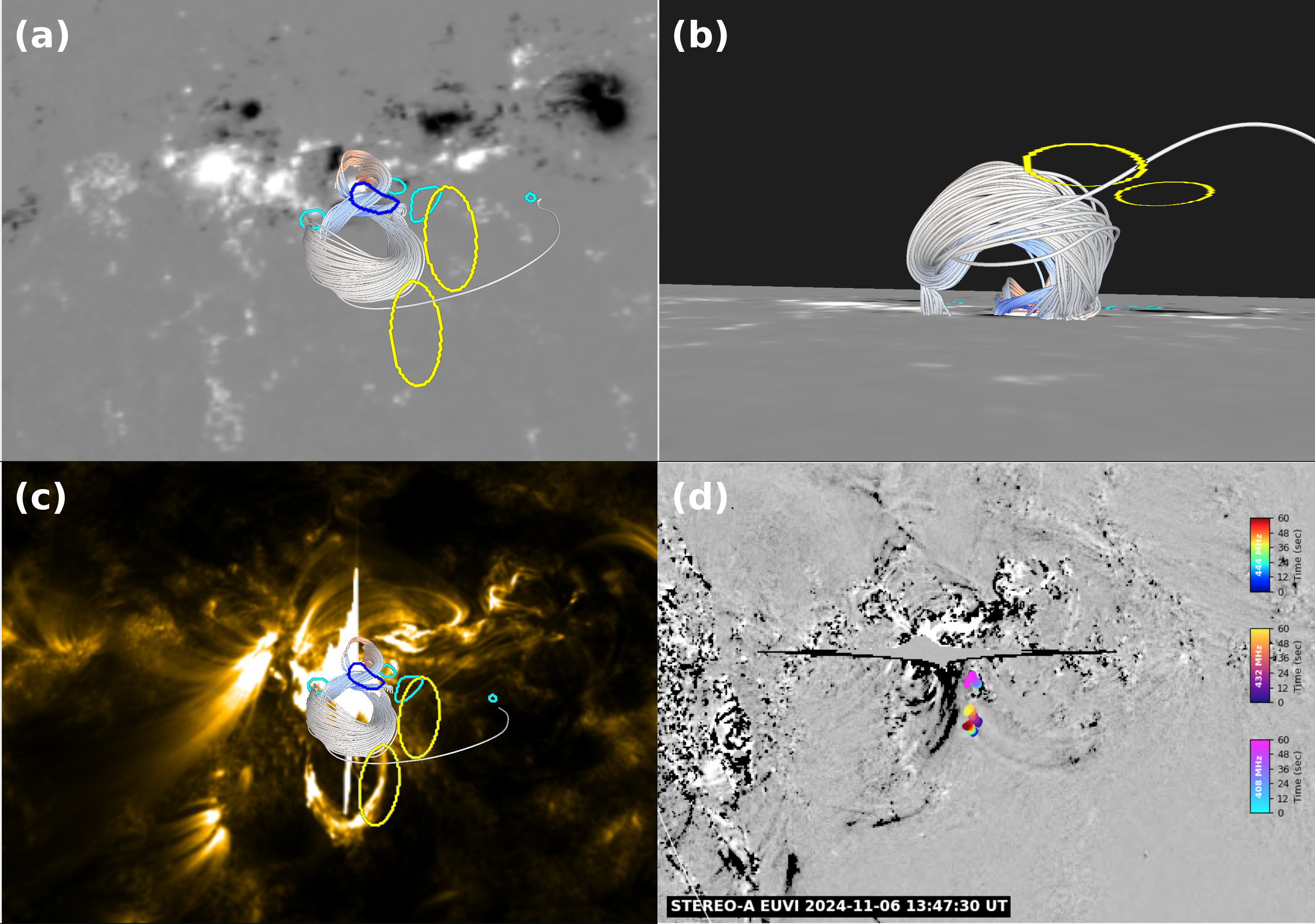} 
    \caption{ The 3D  NLFFF magnetic field configuration of the AR 13883, viewed from top and from the side, respectively. Panels (a) and (b) display the 3D reconstructed coronal field lines overlaid with HXR footpoint (cyan) and looptop (blue) sources at 13:35;00 UT and 13:40:01 UT, along with the 95\% intensity contours of the NRH radio sources (drawn in yellow color) observed at 13:46:53–13:46:58 UT. Panel (c) compares the extrapolated field with SDO/AIA 171 \AA {} EUV loops at 13:40:33 UT, showing good morphological correspondence. Panel (d) represents the reprojected NRH radio sources on STEREO-A 195 \AA {} image at 13:47:30 UT, illustrating their temporal and spatial correlation with the pre-erupting magnetic field structure during the type II burst. }
    \label{Fig:figure7}
\end{figure}


The eruption observed in the EUV images indicates a complex magnetic configuration within the source region. Characterizing the underlying magnetic field is therefore essential for understanding the magnetic configuration of the source region. However, direct measurements of the coronal magnetic field are comparatively rare, given the tenuous plasma conditions of the corona, although off-limb detections \citep{Lin2004, Tomczyk2008, Landi2020, Si2020} and microwave/gyroresonance diagnostics of flaring regions \citep{Fleishman2020, Chen2020} have been reported. Consequently, nonlinear force-free field (NLFFF) extrapolations of photospheric vector magnetograms provide an important means of reconstructing the three-dimensional coronal magnetic field \citep{DeRosa2009, Regnier2013, Wiegelmann2017, Mishra2026}. For the present event, the three-dimensional coronal magnetic topology was reconstructed using an NLFFF extrapolation based on the weighted optimization method of \citet{Wiegelmann2004, Wiegelmann2012}, implemented through the AMaFiL magnetic-field library within the GX Simulator framework \citep{Nita2015, Nita2018, Nita2023}.

The extrapolation was performed using an \texttt{hmi.sharp\_cea\_720s} vector magnetogram \citep{Bobra2014} acquired approximately 12 minutes before the onset of the eruption, providing a snapshot of the pre-eruptive magnetic configuration of the active region. This approach enables us to investigate the magnetic connectivity of the source region and its relationship to the observed X-ray and radio emissions.


To facilitate a direct comparison between the observed emission sources and the reconstructed magnetic field, the STIX hard X-ray and NRH radio source contours were transformed from their observed plane-of-sky coordinates to the cylindrical equal-area (cea) heliographic coordinate system and overlaid on the extrapolated magnetic field. This reprojection minimizes geometric distortions arising from different viewing perspectives and enables a consistent comparison between the observed emission sources and the reconstructed coronal magnetic structure.
Figure \ref{Fig:figure7} presents the reconstructed magnetic configuration of the active region obtained from the NLFFF extrapolation. The extrapolation reveals a magnetic FR rooted along the central polarity inversion line (PIL) and embedded beneath large-scale overlying arcade field lines. The FR extends toward the southeast, consistent with the direction of the observed eruption. Thus, the field lines represent large-scale coronal magnetic connectivity, while the colored contours mark the locations of hard X-ray and radio emission sources relative to the reconstructed magnetic structure. 

Figures \ref{Fig:figure7}(a) and \ref{Fig:figure7}(b) show the top and side views of the extrapolated magnetic field, highlighting the three-dimensional geometry of the FR. The cyan contours correspond to the footpoint locations of the non-thermal hard X-ray emission in the 30–50 keV energy range observed at 13:35:00 UT. These sources are located in the lower solar atmosphere and represent the chromospheric footpoints of flare loops, indicating precipitation of accelerated electrons into the dense chromosphere. The blue contour denotes the looptop hard X-ray source in the same energy range, observed by STIX at 13:40:01 UT. Looptop hard X-ray emission is commonly interpreted as a signature of primary energy release and particle acceleration in the corona. The simultaneous presence of both footpoint and looptop sources suggests that energetic electrons were accelerated in the corona and subsequently propagated along magnetic field lines toward both the chromosphere and the higher corona.

The yellow contours represent the NRH radio sources at the 95\% intensity level observed at 432 and 408 MHz at 13:46:53 UT and 13:46:58 UT, respectively. These radio sources appear a few minutes after the non-thermal hard X-ray emission and are located above the erupting magnetic structure. Their delayed appearance suggests that the radio emission is associated with energetic electrons propagating through the higher corona during the later stages of the eruption. Figure \ref{Fig:figure7}(d) further illustrates the spatial distribution of NRH radio sources relative to the active region projected onto the STEREO-A 195 Å image at 13:47:30 UT. The 99\% intensity contours at 444, 432, and 408 MHz (top, middle, and bottom panels, respectively) exhibit a systematic spatial evolution during the type II burst interval, indicating frequency-dependent source motion consistent with propagation through different coronal heights. The reprojected radio source locations, together with their temporal and spatial evolution, suggest that the emission is closely associated with the evolving erupting structure rather than fixed coronal locations.

The temporal evolution from hard X-ray to radio emission is consistent with an expanding magnetic structure, in which the initial energy release and electron acceleration occur in the low corona, followed by the propagation of energetic electrons and the generation of radio emission at higher coronal heights. The spatial association of the radio sources with the erupting magnetic structure, together with the close correspondence between the eruption radial speed (536 ± 57 km/s) and the shock speed derived from the type II burst (540 ± 39 km/s), supports a scenario in which the erupting FR generated a coronal shock that accelerated electrons responsible for the observed radio emission.

It is important to note that the extrapolated magnetic field represents the pre-eruptive magnetic configuration approximately 12 minutes before the onset of the eruption. Consequently, the field lines shown in Figure \ref{Fig:figure7} do not correspond to the exact magnetic configuration at the times of the hard X-ray and radio observations. The overlaid hard X-ray and radio contours are therefore intended to illustrate the spatial association of the emission sources with the pre-eruptive magnetic topology rather than their exact magnetic connectivity during the event. Furthermore, since both the radio and hard X-ray sources are observed in the plane of the sky, projection effects may introduce apparent offsets between their observed positions and their true three-dimensional locations in the corona.

Nevertheless, the extrapolated magnetic field closely matches the observed EUV coronal structures shown in Figure \ref{Fig:figure7}(c), supporting the reliability of the reconstructed magnetic configuration and indicating that the eruption originated from this pre-existing FR system. Overall, the extrapolated magnetic topology indicates that magnetic energy was stored within a compact FR system prior to eruption. As the magnetic configuration evolved, the system became unstable and released its stored magnetic energy, leading to the eruption of the FR. Although no clear CME was detected in coronagraph observations, the eruption was sufficiently energetic to generate a large-scale coronal disturbance that likely steepened into a shock, producing the observed high-frequency type II radio burst. Therefore, the combined magnetic extrapolation, stereoscopic EUV observations, X-ray diagnostics, and radio measurements consistently support a scenario in which the observed shock was driven by the eruption of a compact magnetic FR rather than by a large-scale white-light CME.

\section{Conclusion} \label{sec:conclusion}
We present a comprehensive multiwavelength investigation of the X2.3-class flare on 6 November 2024 using EUV imaging, radio dynamic spectra, radio imaging, X-ray observations, and NLFFF magnetic field extrapolation to investigate the origin of the associated high-frequency type II radio burst. The main findings are summarized below.

\begin{enumerate}

\item {The magnetic field extrapolation reveals a highly twisted and sheared pre-eruptive magnetic FR capable of storing significant magnetic energy. EUV observations show that this compact FR subsequently erupted during the impulsive phase of the flare. Despite the intense X-class flare magnitude, no corresponding large-scale white-light CME was detected in coronagraph observations, suggesting that the eruption did not develop into a detectable large-scale white-light CME, unlike many metric type II events associated with large-scale CME eruptions.}

\item 
 The expanding compact FR was accompanied by a localized EUV disturbance whose morphology and propagation are consistent with a fast-mode MHD wave in the low corona. The split-band structure of the type II burst yields a density compression ratio corresponding to an Alfvén Mach number of approximately 1.44, consistent with a weak super-Alfvénic shock.

\item 
 The close temporal association between the EUV disturbance and the high-frequency type II burst, together with the agreement between the radio-derived shock speed and the independently measured FR speed, is consistent with a low-coronal shock driven by the expanding compact FR. The spatial association of the hard X-ray and radio sources with the erupting magnetic structure further supports this interpretation.

The present observations suggest that although CME-driven shocks remain the dominant mechanism for generating type II radio bursts, localized expanding FR can also efficiently generate low-coronal shocks, thereby extending our understanding of the origin of high-frequency type II radio bursts.
\end{enumerate}

\begin{acknowledgments}
DP, VVS, and DM acknowledge financial support through the PRL Institute fellowship. The computational resources used in this study were provided by the Param Vikram facility at the PRL. DP is thankful to Dr Manju Sudhakar for discussions regarding HEL1OS data. AK acknowledges support from the ANRF Prime Minister Early Career Research Grant (PM ECRG) program. This work used observations from multiple space- and ground-based instruments. We acknowledge the use of data from the Solar Dynamics Observatory (SDO), including the Helioseismic and Magnetic Imager (HMI) and the Atmospheric Imaging Assembly (AIA), the Solar TErrestrial RElations Observatory (STEREO), the Solar and Heliospheric Observatory (SOHO), the High Energy L1 Orbiting X-ray Spectrometer (HEL1OS) on board Aditya-L1, and the Spectrometer Telescope for Imaging X-rays (STIX) on board Solar Orbiter. HMI magnetogram data were obtained through the Joint Science Operations Center (JSOC), and Geostationary Operational Environmental Satellite (GOES-16) observations were used for flare identification and characterization. We gratefully acknowledge the radio solar database maintained by LESIA and the USN at the Observatoire de Paris for from providing access to the Nançay Radioheliograph and ORFEES data. We also thank the e-CALLISTO network for providing continuous radio spectroscopic observations. 

\end{acknowledgments}

\begin{contribution}
AK conceptualized the project. DP performed the data analysis and wrote the majority of the manuscript. VVS, DM, PD, and NK contributed to the scientific discussions and reviewed the manuscript. All authors contributed to the discussions and summary in the manuscript. All authors reviewed and approved the final manuscript. 
\end{contribution}

\software{}
Sunpy, Matplotlib, Scipy, numpy, IDL, pandas.

\bibliographystyle{aasjournalv7}
\bibliography{Reference}

\end{document}